\documentclass[twocolumn,showpacs,prl,amsmath,amssymb]{revtex4}
\usepackage{graphicx} 
\usepackage{bm}       
\usepackage{hyperref}

\begin{document}

\title{On the effects of geographical constraints on task execution in
  complex networks}

\author{Andr\'e Franceschi de Angelis}\email{andre@ceset.unicamp.br}
\affiliation{ Centro Superior de Educa\c{c}\~ao Tecnol\'ogica,
  Universidade Estadual de Campinas, Rua Paschoal Marmo, 1888, CEP 13484-370,
  Limeira, S\~{a}o Paulo, Brazil }

\author{Gonzalo Travieso}\email{gonzalo@ifsc.usp.br}
\author{Carlos Ant\^onio Ruggiero}\email{toto@ifsc.usp.br}
\author{Luciano da Fontoura Costa}\email{luciano@ifsc.usp.br}
\affiliation{
  Instituto de F\'{\i}sica de S\~{a}o Carlos, Universidade de S\~{a}o Paulo,
  Av.\ do Trabalhador S\~{a}o-carlense 400, Caixa Postal 369, CEP 13560-970,
  S\~{a}o Carlos, S\~{a}o Paulo, Brazil
}

\begin{abstract}
  In the present work we investigate the effects of spatial
  constraints on the efficiency of task execution in systems underlain
  by geographical complex networks where the probability of connection
  decreases with the distance between the nodes.  The investigation
  considers several configurations of the parameters defining the
  network connectivity, and the Barab\'asi-Albert network model is
  also considered for comparisons.  The results show that the effect
  of connectivity is significant only for shorter tasks, that the
  locality of connections implied by the spatial constraints reduces
  efficency, and that the addition of edges can improve the efficiency
  of the execution, although with increasing locality of the
  connections the improvement is small.
\end{abstract}


\maketitle

Great part of the current theoretical and applied research in physics
relies on fast execution of relatively complex algorithms.  Several
problems of current interest can only be solved by using parallel or
distributed computing systems.  A recent trend, namely \emph{grid
  computing} \cite{FosterNature}, often allows more cost-effective
solutions to such problems, involving the combined use of several
general purpose machines. Such an architecture promotes the natural
scaling of the number of computing elements in terms of their
availability and importance of specific problems (more elements can be
assigned to more important problems). The interconnection of machines
participating in grid computing systems involves not only local area
networks, but mainly the Internet. The connectivity in grid computing
is intrinsically dynamic because of its own nature, i.e.\ the fact
that the availability of machines varies with time. In addition, as
the interconnections between such machines are often implemented
through the Internet, the connectivity of the latter inherently
determines the grid architecture \cite{132}. Although grid computing
is inherently more flexible and scalable than traditional parallel and
distributed systems the interconnectivity of the processing elements
in a grid system, combined with specific properties of those elements,
is fundamental for achieving efficiency \cite{A}.  Therefore, given a
specific problem, one needs to select a suitable interconnectivity
which, in the case of Internet-based grid computing, is inherently
constrained by the Internet.

Instead of implementing a specific problem and inferring the
respective performance, a more reasonable and effective means is to
model and simulate the execution of the given algorithms, which is not
only cheaper but can be used to provide additional insights about the
effect of modifying the interconnection and algorithms implementation.
Because of their flexibility for representing almost any discrete
structure, \emph{complex networks}
\cite{Albert02,Dorogovtsev02,B,Costa_surv:2007} represent a natural
resource for modeling distributed computing systems, where the
processing elements are denoted by nodes and their respective
interconnections are represented by edges.  Such a potential is
especially relevant in the case of grid computing, which often
involves interconnectivities substantially different from the highly
regular connections traditionally adopted in parallel computing (e.g.,
meshes, tori, hypercubes, etc.). As a matter of fact, the underlying
Internet is itself a complex network \cite{Faloutsos}.

This work focuses on the problem of executing independent tasks in
diversely interconnected grid computing systems underlain by Internet
connectivity.  A master node partitions the problem to be computed in
independent tasks and distribute the tasks among the other nodes.
Such an investigation extends and complements a previous work
\cite{A}, which applied complex networks in order to investigate the
effect of diverse interconnectivities, including uniform random and
scale free models, on the efficiency of grid computing.  In
particular, in the current work we consider a new interconnectivity
model (growing geometric model, GGM, as well as its extension to
include redundant edges, GGM-RE-n), so as to allow the quantification
of the effects of geographically oriented connectivity and redundant
edges on the performance of grid computing.  This is important because
geographical constraints are fundamental in the Internet.  The
Barab\'asi-Albert network model is also considered for comparison
purposes.

GGM is a novel complex network model which takes into account the
influence of the position of the nodes during network growth.  Once
the geometry of the underlying space is defined, no additional
parameters are required in order to define the GGM evolution.  Instead
of assuming a pre-specified spatial distribution of all nodes (as in
previous geometrical complex networks models), the GGM model involves
the progressive incorporation of new nodes, which are connected to the
closest existing nodes.  Such a dynamics is expected to emulate, to
some accuracy level, the historical evolution of the Internet in
developing regions where, starting from an initial point, the Internet
is progressively extended through the addition of new points connected
to the closest existing outlet.  More specifically, the GGM model
involves the following steps: (i)~The underlying spatial region is
defined (a unit two-dimensional square is adopted in the present
work); (ii)~one of its points is chosen as the initial node; and
(iii)~a new point is randomly (uniformly, in the present work) chosen
within the underlying region and connected to the closest existing
node; (iv)~the previous step is repeated until the desired number of
nodes is reached.  An immediate consequence of such a growth dynamics
is that the resulting network always contains a single connected
component.  More precisely, the obtained network is a tree devoid of
cycles, self-connections and isolated nodes.  In order to allow for
the presence of cycles, which are found in the Internet, especially at
later developmental stages, we also considered an extension of the GGM
model incorporating redundant edges, henceforth called GGM-RE.  By
\emph{redundant} it is meant that the additional edges will not change
the accessibility to any given node, though they will change other
properties such as the node degree, shortest path length and
clustering coefficient.  Note that the inclusion of these redundant
edges will necessarily create cycles in the network.  We will express
the fact that a model has an $n\%$ increase in the number of edges
with respect to the GGM model by the abbreviation GGM-RE-n, such that
GGM-RE-0 corresponds to the initial model (GGM).  Note that, although
more flexible schemes would be possible, in the present work the
redundant edges are incorporated only after the GGM growth is
completed.  The redundant edges are established by linking randomly
chosen pairs of unconnected nodes.  The pairs are chosen with the
following procedure: first a node $i$ is randomly selected; a
different node $j$ is selected, and is connected with $i$ with
probability proportional to $ k_je^{-\alpha d_{ij}}$, where $k_j$ is
the degree of node $j$, $d_{ij}$ is the geographical distance between
$i$ and $j$, and $\alpha$ is the \emph{locality constant} (its value
determines the importance of geographical distance in the
establishment of connections).  The process is repeated until the
desired number of additional edges is achieved.  In the present work,
GGM-RE-n networks were generated through the addition of $10\%$ to
$100\%$ of redundant edges to the same initial GGM network.  In order
to provide a reference for comparisons, the Barab\'asi-Albert
model~\cite{Barabasi97} (BA) was also considered with the same number
of nodes and average degree.

The simulations consider that the initial problem has been partitioned
into $M$ independent tasks, which are distributed by the master
processing element amongst the slaves (i.e.\ each of the other
processing elements represented by the $N-1$ remaining nodes, where
$N$ is the number of nodes in the network), which execute the tasks in
parallel.  All tasks take the same amount of time $L$ to conclude.
The master processing element does not execute tasks itself, but
coordinates their distribution and collects the results.  It is
assumed that the master can perform unimpeded by delays or
bottlenecks.  The slave processing elements process each task
independently, i.e.\ without the need to communicate with other
slaves.  Because the number of tasks is not necessarily equal to an
integer multiple of the number of slave processors, idle processing
elements may be found at any time during the overall execution.  By
similar reasoning, some slaves may never receive a task in case the
number of slaves is larger than the number of tasks.  The simulations
assume a communication cost proportional to the total number of links
between the master and each of the slaves, with the same communication
times for all links; the time taken for the communication through one
link is used as the unity of time for the simulation.  As no
additional costs (e.g., due to routing or congestion) are considered,
the performance of the overall execution is directly affected by the
network topology.  For instance, the shortest the distance between the
nodes, the better the performance.  At the same time, the selection of
the node used as master can significantly influence the overall
performance.  In case a poorly connected node is chosen for that role,
higher communication costs will be implied.  In order to average over
such effects, the simulation algorithm considers each of the nodes as
master for each generated network.  The parallel execution time $T_P$
is defined as the time taken from the dispatch of the first task to a
slave until the arrival of the result from the last task in the master
node.  The parallel \emph{speedup} ($S$) is the ratio between the time
that would be taken for the sequential execution ($M L$ in the
considered application model) and the parallel execution time, $S = M
L / T_P$.  The performance of each simulation is quantified in terms
of its \emph{efficiency} $E$, defined as the speedup divided by the
number of processing elements, $E=S/N$.  The values reported are
averages $\langle E \rangle$ of the efficiency over 30~networks for
each model and parameter values.

\begin{figure}
  \includegraphics[width=0.45\textwidth]{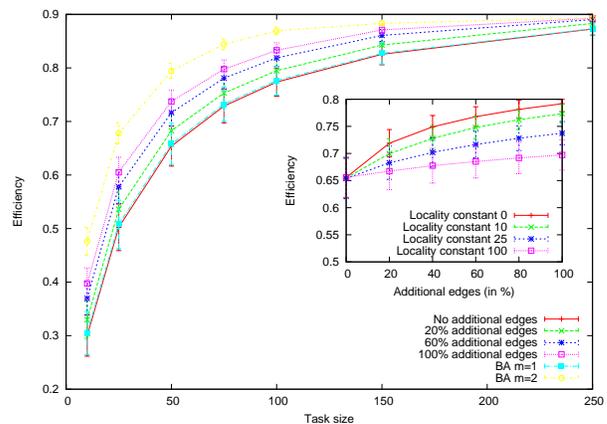}
  \caption{(Color online) Parallel average efficiency $\langle E
    \rangle$ as a function of the size $L$ of tasks for the GGM-RE
    model with locality constant $25$, and $0$ to $100\%$ additional
    edges, compared the BA model with $m=1$ and $m=2$. The inset shows
    the effect of additional edges on the efficiency, for different
    values of the locality constant and $L=50$.}
  \label{fig:edges}
\end{figure}

\begin{figure}
  \includegraphics[width=0.45\textwidth]{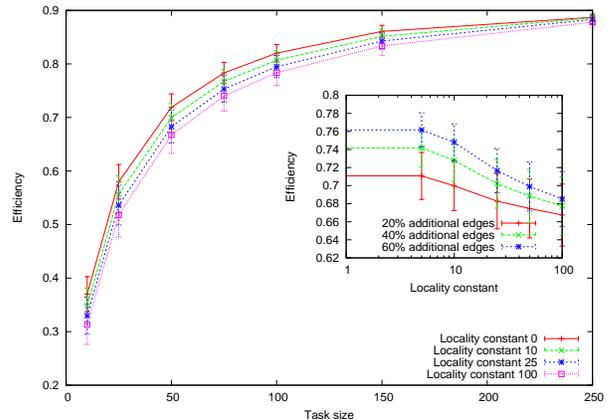}
  \caption{(Color online) Parallel average efficiency $\langle E
    \rangle$ as a function of the size $L$ of tasks for the GGM-RE
    model with $20\%$ additional edges and locality constant from $0$
    to $100$.  The inset shows the effect of the locality constant on
    the efficiency, for different connectivities and $L=50$.}
  \label{fig:const}
\end{figure}

Networks of $N=1000$ nodes were generated according to the above
described models; all results are averages of $30$ realizations.
Figure~\ref{fig:edges} shows the effect of the size of the tasks $L$
on the efficiency, for fixed $M=10\,000$.  It is clear that, for
sufficiently large tasks, the network topology is not an important
factor in the efficiency, with significant differences only for $L$
smaller than $100$.  Network topology influences the processing
efficiency through communication costs; if the computation time of a
task is of the order of the time taken to send it to the client and
than back to the master, communication cost becomes an important
factor; for large $L$ (about two orders of magnitude larger than the
communication latency), the communication cost, and therefore the
network topology, has only small influence in the efficiency.  The
figure also shows the effect (specially for smaller values of $L$) of
increased connectivity: additional edges shifts the efficiency curve
up.  But the effect on efficiency of adding new edges is decreasing;
this can be better seen in the inset, where average efficiency versus
additional edges is plotted for $L=50$.  Comparing the GGM-RE model
without additional edges with the BA model with $m=1$ we see that the
efficiencies are almost indistinguishable.  This result agrees with
previous results \cite{A}, which found that the efficiency is strongly
influenced by the average degree and the fraction of nodes in the
largest connected component.  In the present case, both networks have
the same average degree and are totally connected.  The same is not
true for GGM-RE with $100\%$ additional edges and BA with $m=2$
(again, both with the same average degree), where the efficiency of
the BA model is clearly superior.  This is due to the emphasis on
connecting spatially close nodes in the geographical model: this
reduces the amount of shortcuts between different regions of the
network, increasing average distances between nodes.  This effect is
more pronounced as the locality constant is increased, as can be seen
in the inset: networks with large locality constants almost do not
profit from new edges.  The effect of geographical locality in the
construction of the network can also be seen in
Figure~\ref{fig:const}, which shows the efficiency as a function of
task size for different values of the locality constant.  An higher
locality constant is associated with a lower efficiency, as the number
of connections among geographically distant regions tend to decrease,
increasing the average distances between nodes and therefore the
communication times.  The inset in Figure~\ref{fig:const} plots
efficiency agains the locality constant, and shows that small locality
constants have little effect, though for values grater than $5$ the
efficiencies decrease (note that the range of efficiency values
spanned is small).

The results presented in this letter show that spatial restrictions,
which influence the development of connectivity in complex networks,
are a limiting factor for the efficiency distributed task execution in
such networks.  Because of the lower probability of creation of
shortcuts among distant sets of nodes, the efficiency of information
transfer in the network is reduced.  With the aim of achieving
efficient execution of tasks in complex networks it is therefore
important to promote the formation of long-range connections during
the network contruction process.

\begin{acknowledgments}

Luciano da F. Costa thanks CNPq (308231/03-1) and FAPESP (05/00587-5)
for sponsorship.  The authors thank FAPESP for financial support
(03/08269-7).

\end{acknowledgments}

\bibliography{geogrid}

\end{document}